\documentstyle[preprint,aps]{revtex}

\begin{document}
\draft
\title{Quantum open systems approach to current noise in resonant tunnelling
junctions.}
\author{He Bi Sun and G.J.Milburn}
\address{Department of Physics,\\
University of Queensland,\\
St Lucia 4072, Australia}
\date{\today}
\maketitle

\begin{abstract}
A quantum Markovian master equation is derived to describe the current noise
in resonant tunnelling devices. This equation includes both incoherent and
coherent quantum tunnelling processes. We show how to obtain the population
master equation by adiabatic elimination of quantum coherences in the
presence of elastic scattering. We calculate the noise spectrum for a double
well device and predict sub-shot noise statistics for strong tunnelling
between the wells. The method is an alternative to Greeen's functions
methods, and population master equations for very small coherently coupled
quantum dots.
\end{abstract}


\section{Introduction}

Quantum features of conductance in mesoscopic electronics is currently a
major theoretical and experimental research interest in condensed matter
physics\cite{mesoscopics}. Developments are driven by two complementary
imperatives. Firstly a technological trend to fabricate devices on smaller
and smaller scales is rapidly approaching the point where quantum effects
will become a problem unless explicit attempts to exploit quantum features
are made. Quantum tunnelling can lead to undesired coupling between
fabricated structures. On the other hand, tunnelling offers the possibility
of very fast switching times. Secondly, the new devices require improvements
to the theoretical description of electronic transport in a low temperature,
high mobility regime. Small devices with very long coherence times can be
dominated by coherent quantum effects. It is becoming increasingly clear
that intrinsic quantum fluctuations play an important role at low
temperatures\cite{Mohanty1997}

Current noise in resonant tunnelling devices (RTD) provides a path to
understanding noise in the deep quantum domain. In a biased RTD one or more
bound quantum states are coupled incoherently to two electron reservoirs
maintained at different chemical potential. There are a number of
experimental\cite{Li90,vdRoer91,Brown92,Liu95} and theoretical\cite
{Landauer95,Frensley90,Chen91,Davies92,Hershfield93,Liu96} results. RTDs
involve exchange of fermions between the reservoirs, and the bound states.
We propose in this paper an approach to such devices based on quantum markov
master equations\cite{Gardiner}. Such an approach to quantum noise in
nonequlibrium systems has been used with great success in quantum optics.
This provides an alternative approach to the conventional Green's functions
methods, and offers additional physical insights. For example it enables one
to deal with coherent coupling between adjacent well states which couples
off-diagonal elements of the density matrix in the occupation number basis
and cannot be described by population master equations. Such coupling can
occur in strongly coupled quantum dots, as in the recent experiments of
Blick et al\cite{Blick98} and Oosterkamp et al\cite{Oosterkamp98}.

If the strength of this coherent coupling dominates the time scales of
elastic and inelastic relaxation, a population master equation cannot
describe the system. Coherently coupled Nanostructures are likely to become
increasingly important and thus there is considerable motivation to develop
theoretical schemes that go beyond population rate equations.

In the first part of this paper we derive the operator master equation to
describe a bound electronic system coupled incoherently to two reservoirs.
We then apply this equation to calculate the current two-time correlation
function for a single well, with a single bound state. This model of course
can equally well be treated by a population master equation approach as in
the approach of Carlos Egues et al.\cite{Egues94}, but we rederive the known
results simply to display the method in a familiar context. In section III
we apply our methods to treat the case of coherent coupling between the
bound states of adjacent wells. In this case our approach yields results
that go beyond the traditional population master equation approach. We
derive the current spectrum in the device and demonstrate new features that
arise precisely because of the coherent coupling between the two wells. To
make contact with previous work we show that in the limit when elastic
scattering dominates the coherent coupling, a population rate equation may
be derived that is equivalent, in the appropriate limit, to that obtained by
Carlos Egues et al.\cite{Egues94}.

\section{The master equation}

We begin with the derivation of the master equation for a single quantum
tunnelling channel connecting two reservoirs under external bias. This
system is quite adequately described by other methods, including population
master equations. However we treat it here simply to demonstrate our
approach in a familiar setting. Therefore our results are not new and could
equally well be obtained by other methods. This is not the case for the
coherently coupled double well system we discuss next. The hamiltonian
describing this process is given by\cite{Chen91} 
\begin{eqnarray}
H &=&\sum_{k}\varepsilon _{k}^{E}a_{k}^{\dagger }a_{k}+\varepsilon
_{c}c^{\dagger }c+\sum_{p}\varepsilon _{p}^{C}b_{p}^{\dagger }b_{p} \\
&&\mbox{}+\sum_{k}(T_{Ek}c^{\dagger }a_{k}+T_{Ek}^{*}a_{k}^{\dagger }c) 
\nonumber \\
&&\mbox{}+\sum_{p}(T_{Cp}b_{p}^{\dagger }c+T_{Cp}^{*}c^{\dagger }b_{p}) 
\nonumber
\end{eqnarray}
where $a_{k}(a_{k}^{\dagger }),c(c^{\dagger })$ and $b_{p}(b_{p}^{\dagger })$
are the annihilation (creation) operators of electrons in the emitter ($E$)
reservoir, in the central quantum well and in the collector ($C$) reservoir
respectively. The energy of the bound state without bias is $\varepsilon
_{0} $ which under bias becomes $\varepsilon _{c}=\varepsilon _{0}-\alpha eV$
where $\alpha $ is a structure dependent coefficient. The single particle
energies in the emitter and collector are respectively, $\varepsilon
_{k}^{E}=k^{2}/2m$ and $\varepsilon _{p}^{C}=p^{2}/2m-eV$. The energy
reference is at the bottom of the conduction band of the emitter reservoir.

The fourth and fifth terms in the Hamiltonian describe the coupling between
the quasibound electrons in the well and the electrons in the reservoir. The
tunnelling coefficients $T_{Ek},T_{Cp}$ depend on the barrier profile and
the bias voltage. We will assume that at all times the two reservoirs remain
in thermal equilibrium, with chemical potentials $\mu_C,\mu_E$, with $%
\mu_E-\mu_C=eV$, despite the tunnelling of electrons. This is one of the key
defining characteristics of a reservoir. It assumes in effect that two very
different time scales describe the dynamics of the reservoirs and the
quaisbound quantum state in the well.

In the interaction picture the Hamiltonian may be written as 
\begin{equation}
H_{I}(t)=\hbar \sum_{i=1}^{2}(c^{\dagger }\Gamma _{i}(t)e^{i\omega
_{0}t}+c\Gamma _{i}^{\dagger }(t)e^{-i\omega _{0}t})  \label{ham}
\end{equation}
where the bound state frequency is $\omega _{0}=\varepsilon _{c}/\hbar $ and
the reservoir operators are given by 
\begin{eqnarray}
\Gamma _{1}(t) &=&\sum_{k}T_{Ek}a_{k}e^{-i\omega _{k}^{E}t} \\
\Gamma _{2}(t) &=&\sum_{p}T_{Cp}b_{p}e^{-i\omega _{p}^{C}t}
\end{eqnarray}
where 
\begin{eqnarray}
\omega _{k}^{E} &=&\frac{\varepsilon _{k}^{E}}{\hbar } \\
\omega _{p}^{C} &=&\frac{\varepsilon _{p}^{C}}{\hbar }
\end{eqnarray}

We now obtain an equation of motion for the density operator of the bound
state, $\rho (t)$, in the well following the standard method based on second
order perturbation theory and tracing over reservoir states\cite{WallsMilb94}%
. Thus we need 
\begin{equation}
\frac{d\rho (t)}{dt}=-\frac{1}{\hbar ^{2}}%
\int_{0}^{t}dt_{1}Tr_{R}[H_{I}(t),[H_{I}(t_{1}),\rho _{R}\otimes \rho (t)]]
\end{equation}
where $\rho _{R}$ is the thermal equilibrium state of the two reservoirs,
and $Tr_{R}$ denotes a trace over the reservoir variables. Note that the
factorisation of the well state and reservoir states has been assumed. This
is reasonable if the well state and the reservoir states are initially
uncorrelated and provided there is a wide separation in the relaxation time
scales of the well state and the reservoirs. The only non zero correlation
functions we need to compute are 
\begin{eqnarray}
I_{E1}(t) &=&\int_{0}^{t}dt_{1}\langle \Gamma _{1}^{\dagger }(t_{1})\Gamma
_{1}(t_{1})\rangle e^{-i\omega _{0}(t-t_{1})} \\
I_{E2}(t) &=&\int_{0}^{t}dt_{1}\langle \Gamma _{1}(t_{1})\Gamma
_{1}^{\dagger }(t_{1})\rangle e^{-i\omega _{0}(t-t_{1})} \\
I_{C1}(t) &=&\int_{0}^{t}dt_{1}\langle \Gamma _{2}^{\dagger }(t_{1})\Gamma
_{2}(t_{1})\rangle e^{-i\omega _{0}(t-t_{1})} \\
I_{C2}(t) &=&\int_{0}^{t}dt_{1}\langle \Gamma _{2}(t_{1})\Gamma
_{2}^{\dagger }(t_{1})\rangle e^{-i\omega _{0}(t-t_{1})}
\end{eqnarray}
In order to illustrate the important physical approximations required in
deriving the master equation we will now explicitly evaluate the first of
these correlation functions.

Using the definition of the reservoir operators and the assumed thermal
Fermi distribution of the electrons in the emitter we find 
\begin{equation}
I_{E1}(t)=\sum_{k}\bar{n}_{Ek}|T_{Ek}|^{2}\int_{0}^{t}dt_{1}e^{(i(\omega
_{k}^{E}-\omega _{0})t}
\end{equation}
As the reservoir is a large system by definition we can replace the sum over 
$k$ by an integral to obtain 
\begin{equation}
I_{E1}(t)=\int_{0}^{\infty }\frac{d\omega }{2\pi }\rho (\omega )\bar{n}%
(\omega )|T_{E}(\omega )|^{2}\int_{0}^{\tau }d\tau e^{i(\omega -\omega
_{0})\tau }
\end{equation}
where we have changed the variable of time integration. The dominant term in
the frequency integration will come from frequencies near $\omega _{0}$ as
the time integration is significant at that point. We assume that the bias
is such that the quasibound state is well below the Fermi level in the
emitter. This implies that near $\omega =\omega _{0}$, the average
occupation of the reservoir state is very close to unity. This is an
effective low temperature approximation. Now we make the first Markov
approximation. We assume that the function $\rho (\omega )\bar{n}(\omega
)|T_{E}(\omega )|^{2}$ is slowly varying around $\omega =\omega _{0}$, and
thus the frequency integration will lead to a function which is a rapidly
decaying function of time compared to dynamical time scales for the
quasibound state. This implies that on time scales of interest in an
experiment we can extend the upper limit of the time integration to infinity
as a good approximation. In that case $I_{E1}$ becomes time independent and
may be approximated by 
\begin{eqnarray}
I_{E1}(t) &\approx &\int_{0}^{\infty }\frac{d\omega }{2}\rho (\omega
)|T_{E}|^{2}\delta (\omega _{0})  \nonumber \\
&=&\gamma _{L}(\omega _{0})
\end{eqnarray}
which defines the effective rate $\gamma _{L}$ of injection of electrons
from the left reservoir (the emitter) into the quasibound state of the well.
This rate will have a complicated dependence on the bias voltage through
both $\omega _{0}$ and the coupling coefficients $|T_{E}(\omega )|$. In this
paper we do not address this issue. We simply seek the noise properties as a
function of the rate constants.

Evaluating all the other correlation functions under similar assumptions we
find that the quantum master equation for the density operator representing
the well-state in interaction picture is given by 
\begin{eqnarray}
\frac{d\rho }{dt} &=&{\cal L}\rho  \nonumber \\
&=&\frac{\gamma _{L}}{2}(2c^{\dagger }\rho c-cc^{\dagger }\rho -\rho
cc^{\dagger })  \nonumber \\
&&\mbox{}+\frac{\gamma _{R}}{2}(2c\rho c^{\dagger }-c^{\dagger }c\rho -\rho
c^{\dagger }c),  \label{master1}
\end{eqnarray}
$\gamma _{L}$ and $\gamma _{R}$ are constants determining the rate of
injection of electrons from emitter into the well and from the well into the
collector respectively. The rate constants can be determined by a
self-consistent band calculation involving the bias voltage. Two Poisson
processes shown in the master equations: the injection of electrons into the
well described by the first term in the right hand and the emission of
electrons out of the well by the second term, are conditioned by the rates $%
\gamma _{L}\langle cc^{\dagger }\rangle (t)$ and $\gamma _{R}\langle
c^{\dagger }c\rangle (t)$:

\begin{equation}
E\left( dN_{E}(t)\right) =\gamma _{L}\langle cc^{\dagger }\rangle dt;E\left(
dN_{C}(t)\right) =\gamma _{R}\langle c^{\dagger }c\rangle dt  \label{poisson}
\end{equation}
where the average is taken with respect to the well state at any time t. The
master equation \ref{master1} is diagonal in the occupation number
representation. The mean occupation number $\overline{n}=Tr\left( c^{\dagger
}c\rho (t)\right) $can therefore be determined easily from the rate equation

\begin{equation}
{\displaystyle {d\overline{n} \over dt}}%
=\gamma _{L}(1-\overline{n})-\gamma _{R}\overline{n}
\end{equation}
However the occupation number of the well states is not directly measured in
current experiments. The current noise is a fluctuation in classical
stochastic processes. It is measured in the relatively high temperature
reservoirs of the leads, well away from the well state, and the strong, fast
electron-electron interactions in the reservoir establish the classical
level of the observed variable. It is however conditioned on the underlying
quantum stochastic process in the well, which is described by the master
equation. We thus have the familiar problem of connecting the observed
classical stochastic process to the quantum source of information in an open
quantum system. In this problem we proceed as follows. The current pulse in
the emitter and collector may be determined from the Ramo-Shockley theorem 
\cite{R-S1939}. For a symmetric geometry this takes the form,

\begin{equation}
i(t)dt=%
{\displaystyle {e \over 2}}%
\left( dN_{E}(t)+dN_{C}(t)\right)
\end{equation}
The connection to the quantum source is then made by equations (\ref{poisson}%
). Using Eq(16), the average current is given by $E(i(t))=\gamma _{L}(1-%
\overline{n})-\gamma _{R}\overline{n}$. In the steady state this is $%
i_{\infty }=\frac{e\gamma _{L}\gamma _{R}}{\gamma }\label{sscurrent}$ where $%
\gamma =\gamma _{L}+\gamma _{R}$ and the subscript $\infty $ indicates the
steady state.

The fluctuations in the observed current, $i(t)$ are quantified by the
two-time correlation function: 
\begin{equation}
G(\tau )=\frac{e}{2}i_{\infty }\delta (\tau )+\left\langle I(t),I(t+\tau
)\right\rangle _{\infty }^{\tau \neq 0}  \label{Eq-corr}
\end{equation}
To relate these classical averages to the fundamental quantum processes
occurring in the well we apply the theory of open quantum system \cite
{Carmicael1993} to the present system and calculate the following
correlation components with $\tau >0$\cite{Wiseman1993}:

\begin{equation}
E\left( dN_{E}(t+\tau )dN_{E}(t)\right) =\gamma _{L}^{2}Tr\left( cc^{\dagger
}e^{{\cal L}\tau }c^{\dagger }\rho _{\infty}c\right) dt^{2}
\end{equation}

\begin{equation}
E\left( dN_{C}(t+\tau )dN_{C}(t)\right) =\gamma _{R}^{2}Tr\left( c^{\dagger
}ce^{{\cal L}\tau }c\rho _{\infty}c^{\dagger }\right) dt^{2}
\end{equation}

\begin{equation}
E\left( dN_{E}(t+\tau )dN_C(t)\right) =\gamma _{L}\gamma _{R}Tr\left(
cc^{\dagger }e^{{\cal L}\tau }c\rho _{\infty}c^{\dagger }\right) dt^{2}
\end{equation}

\begin{equation}
E\left( dN_{C}(t+\tau )dN_{E}(t)\right) =\gamma _{R}\gamma _{L}Tr\left(
c^{\dagger }ce^{{\cal L}\tau }c^{\dagger }\rho _{\infty}c\right) dt^{2}
\end{equation}
Calculating the above correlation components using master equation with
corresponding initial conditions, and substituting them together with the
shot-noise component into equation \ref{Eq-corr} yield: 
\begin{eqnarray}
G(\tau ) = 
{\displaystyle {ei_{\infty } \over 2}}%
\delta (\tau )+%
{\displaystyle {ei_{\infty } \over 4}}%
\left( 1-%
{\displaystyle {4\gamma _{L}\gamma _{R} \over \gamma ^{2}}}%
\right) \gamma e^{-\gamma \left| \tau \right| }
\end{eqnarray}
Thus the spectral density of current fluctuation in frequency domain is
given by

\begin{eqnarray}
S(\omega ) & = & 2\int_{0}^{\infty }G(\tau )\left( e^{i\omega \tau
}+e^{-i\omega\tau }\right)d\tau  \nonumber \\
& = & ei_{\infty }\left( 1+\left( 1-\frac{4\gamma _{L}\gamma _{R}}{\gamma^{2}%
}\right) \frac{\gamma^{2}}{\gamma^{2}+\omega^{2}}\right)
\end{eqnarray}
The current Fano factor $F(\omega )$ is defined as the ratio of current
noise density over the full shot noise density, and for low frequencies ($%
\gamma \gg \omega $):

\begin{equation}
F(0)=%
{\displaystyle {S(0) \over 2ei_{\infty }}}%
=1-%
{\displaystyle {2\gamma _{L}\gamma _{R} \over \gamma ^{2}}}%
\end{equation}
The shot noise is suppressed and reaches the minimum of 50 \% in a symmetric
structure with $\gamma _{L}=\gamma _{R}.$ The result is same as those
derived by Chen and Ting \cite{Chen91} using non-equilibrium Green's
function method. The result can also be obtained by a classical master
equation calculation \cite{Davies92}. However the classical master equation
cannot be used to treat the case of coherent coupling in a double well
system discussed below.

The suppression of fluctuations at low frequency is due to the exclusion
principle in the well state, reflected in the master equation by the
appearance of the anti-commuting field operators. No electron can tunnel
onto the well if an electron is already there. We need to wait a time of the
order of $\gamma ^{-1}$ for the electron to tunnel back out into the
collector. Thus strong anti-correlations are established in the two
fundamental Poisson processes, $dN_{i}(t)$. If the tunnelling particles were
bosons, the well could accumulate a large number of particles, enhancing the
probability for emission into the collector. This would lead to a rapid
bunching of emission events into the collector and a super-shot noise
current would result. At high frequencies, we are looking at fast processes
in which an electron tunnels into the well and immediately tunnels out. The
Fano factor at high frequencies is 0.5 due to the assumed form of
Ramo-Shockley theorem.

\section{Noise properties of a coherently coupled double well structure}

We now apply our approach to a triple barrier and double quantum well
involving elastic scattering within the wells and coherent coupling between
the wells. The main procedures are parallel to those in single well case but
now involve off-diagonal elements of the density matrix. The master equation
is, 
\begin{eqnarray}
\frac{d\rho }{dt} &=&\frac{\gamma _{L}}{2}\left( 2c_{1}^{\dagger }\rho
c_{1}-c_{1}c_{1}^{\dagger }\rho -\rho c_{1}c_{1}^{\dagger }\right)  \nonumber
\\
&&\mbox{}+\frac{\gamma _{R}}{2}\left( 2c_{2}\rho c_{2}^{\dagger
}-c_{2}^{\dagger }c_{2}\rho -\rho c_{2}^{\dagger }c_{2}\right)  \nonumber \\
&&-\eta _{1}\left[ c_{1}^{\dagger }c_{1},\left[ c_{1}^{\dagger }c_{1},\rho
\right] \right]  \nonumber \\
&&-\eta _{2}\left[ c_{2}^{\dagger }c_{2},\left[ c_{2}^{\dagger }c_{2},\rho
\right] \right]  \nonumber \\
&&-i\Omega \left[ (c_{1}^{\dagger }c_{2}+c_{2}^{\dagger }c_{1}),\rho \right]
\end{eqnarray}
where $c_{1}(c_{1}^{\dagger })$ and $c_{2}(c_{2}^{\dagger })$ are
annihilation (creation) operator of electron in the left and right quantum
well respectively, $\eta _{i}$ is the rate of elastic scattering
(electron-phonon for example) in the $ith$ well, and $\Omega $ is the
coherent coupling rate between the two well states. The irreversible term
describing the elastic scattering is derived in much the same way as the
inelastic tunnelling terms that describe electrons entering and leaving the
device, with one additional assumption. To get a Markov master equation for
number conserving scattering events we must assume that the temperature of
the bath describing such processes is high enough that the bath states are
well away from the ground states. This is not a very restrictive assumption
for realistic devices at milliKelvin temperatures. The deviations that can
result for very low temperatures are described in Gardiner\cite{Gardiner}
The derivation of the scattering term in the master equation (27) is
detailed in the appendix.

The last term in this equation represents a coherent coupling between the
two wells and causes a single electron to periodically tunnel backward and
forward between the two wells, until it is eventually lost through the final
barrier. Recently Blick et al\cite{Blick98}, have made measurements on a
structure that can be roughly approximated by our model. As they point out
this device exhibits a new feature, in that a single electron can be in a
superposition state between the two wells and is thus like an artificial
molecule. We first derive the noise features in the presence of this
coherent coupling. We will then show that, in the limit of strong elastic
scattering, $\eta _{i}\ >>\ \Omega $ , the system can be described in terms
of population rate equations that have been extensively used in the past.

The steady state current is easily found to be given by 
\begin{equation}
i_{\infty }=\frac{2e\Omega^{2}\gamma_{e}}{\gamma_{e}^{2}+2\gamma_{e}%
\eta_{e}+4\Omega^{2}}
\end{equation}
for a symmetric system, $\gamma _{L}=\gamma _{R}\equiv \gamma _{e}$; $\eta
_{1}=\eta _{2}\equiv \eta _{e}$ . The appropriate correlation functions may
be evaluated to give, 
\begin{eqnarray}
\left\langle I(t),I(t+\tau )\right\rangle_{\infty} & = & \left(\frac{%
e\gamma_{e}\Omega} {\lambda_{+}\lambda_{-}}\right )^2\left\{
4\Omega^{2}\right .  \nonumber \\
& & \mbox{}+\frac{1}{4\Delta } \left[ f_+e^{\lambda _{+}\tau
}+f_-e^{\lambda_{-}\tau}\left .\right] \right\}
\end{eqnarray}
where 
\begin{equation}
f_\pm=\left( \gamma _{e}-\eta _{e}\pm\Delta \right) \left( \Delta\pm\eta_{e}
\right)\left(\gamma _{e}+\eta _{e}+\Delta \right)
\end{equation}
and $\Delta \equiv \sqrt{\eta _{e}^{2}-4\Omega ^{2}}$; $\lambda
_{\pm}=-\gamma _{e}-\eta _{e}\pm \Delta $. The noise spectra are derived in
two cases. In case1: $\eta _{e}^{2}>4\Omega ^{2}$, when the elastic
scattering rate, $\eta_i$, is higher than the coherent coupling rate between
the well states, the current noise spectrum is 
\begin{eqnarray}
S(\omega ) & = & 2ei_{\infty }\left\{ \frac{1}{2}+\frac{\gamma_{e}}{4\Delta }%
\left[ \frac{(\eta_{e}+\Delta )(\gamma_{e}-\eta _{e}+\Delta )}{%
(-\gamma_{e}-\eta_{e}+\Delta )^{2}+\omega ^{2}}\right .\right .  \nonumber \\
& &\mbox{}+\frac{(-\eta_{e}+\Delta )(\gamma_{e}-\eta_{e}-\Delta)}{%
(-\gamma_{e}-\eta_{e}-\Delta )^{2}+\omega ^{2}}\left .\left .\right] \right\}
\end{eqnarray}

The current Fano factor against normalised frequency is plotted in Fig. 1
where the spectrum shows Lorentzian feature. In case2: $\eta
_{e}^{2}<4\Omega ^{2}$, the opposite situation, when coherent coupling is
much stronger than elastic scattering, the noise spectrum 
\begin{equation}
S(\omega )=2ei_{\infty }\left\{ \frac{1}{2}+\frac{\gamma_e}{\tilde{\Delta}}%
\Im\left [\frac{(\gamma_e-\eta_e+i\tilde{\Delta})(\eta_e+i\tilde{\Delta})}{%
(-\gamma_e-\eta_e+i\tilde{\Delta})^2+\omega^2}\right ]\right \}
\end{equation}
where $\widetilde{\Delta }=\sqrt{4\Omega ^{2}-\eta _{e}^{2}}$, is
symmetrically double peaked about the free particle frequency ($\omega =0$)
as shown in the Fig. 2. A comparison of these two quantum processes is shown
in Fig. 3. When elastic scattering increases, the Fano factor increases.
Increasing the coherent coupling results in noise suppression and the double
peak feature are more significant as the two well coupling and the quantum
correlations are stronger. Further when elastic scattering is extremely
weak: $\eta _{e}\rightarrow 0$, and coherent coupling is strong: $\Omega \gg
\gamma _{e}$, the steady state current $i_{\infty }\rightarrow 
{\displaystyle {e\gamma _{e} \over 2}}%
$ approaches the single well case as expected in this limit. A significant
outcome is that the best noise reduction at low frequency when $\eta _{e}=0$
reaches 0.22.

The coherent tunnelling between the two wells has a strong effect on the
noise characteristics. Electrons are periodically transferred between the
two wells at the tunnelling frequency. If an electron from the emitter is
injected into the first well, no further electrons can enter this well until
this electron is removed, which takes place on a time scale determined by $%
\Omega ^{-1}$. Thus at frequencies smaller than $\Omega $, noise is
suppressed by the exclusion principle, just as for the single well case. At
the tunnel frequency however we expect the noise to increase, as electrons
injected into the first well are quickly cycled to the second well, where
they can incoherently escape to the collector. This explains the two peaked
structure of the noise power spectrum. In the case of large $\Omega $
however, coherent coupling dominates. In that case if an electron tunnels
into the first well it periodically returns to that well at a frequency of $%
2\Omega $. To see this it is sufficient to note that the two levels which
are degenerate in the absence of tunnelling become split into symmetric and
anti symmetric combinations, separated in energy by $2\hbar\Omega$. A state
initially localised in one well can then be written as a linear combination
of the two new eigenstates. The phase difference in the superposition
rotates through $\pi$ at the frequency $\Omega$ which leads to a state
localised in the other well. This is just the standard description of
tunnelling in a two state system. The periodic of return of the electron to
the first well suppresses another electron from entering the well. Thus at
large values of $\Omega $ we expect noise suppression to occur at $\omega
=2\Omega $. This behaviour is indeed seen in figure4.

We now show that in the limit of strong elastic scattering $\eta_i\ >>\
\Omega$ (Case 1 above), a population master equation can be derived that
describes a classical sequential tunnelling structure. The sequential model
is traditionally formulated in terms of a classical master equation for the
occupation probabilities of each well. In our case, we have restricted the
discussion to a single bound state in each well and thus the maximum
population in each well is unity. However we can derive an equivalent
classical master equation to describe sequential tunnelling even in this
case.

Our method is an extension of adiabatic methods used in quantum optics to
obtain rate equations. We assume that the off-diagonal elements of the
double well density operator are rapidly damped due to the elastic
scattering rates $\eta _{i}$. The off-diagonal elements are then assumed to
relax almost instantaneously to their steady state values and adiabatically
follow the more slowly changing diagonal matrix elements.

From Eq(27) we find the following equations of motion for the matrix
elements in the occupation number basis for each well,\newline

$%
{\displaystyle {d \over dt}}%
<n_{1}n_{2}|\rho |m_{1}m_{2}>$

$=\left\{ 
\begin{array}{c}
-%
{\displaystyle {\gamma _{L} \over 2}}%
[(n_{1}+1)\delta _{n_{1},0}+(m_{1}+1)\delta _{m_{1},0}]-%
{\displaystyle {\gamma _{R} \over 2}}%
(n_{2}\delta _{n_{2},1}+m_{2}\delta _{m_{2},1}) \\ 
-\eta _{1}(n_{1}^{2}\delta _{n_{1},1}-2n_{1}m_{1}\delta _{n_{1},1}\delta
_{m_{1},1}+m_{1}^{2}\delta _{m_{1},1}) \\ 
-\eta _{2}(n_{2}^{2}\delta _{n_{2},1}-2n_{2}m_{2}\delta _{n_{2},1}\delta
_{m_{2},1}+m_{2}^{2}\delta _{m_{2},1})
\end{array}
\right\} <n_{1}n_{2}|\rho |m_{1}m_{2}>$

$+\gamma _{L}\delta _{n_{1},1}\delta _{m_{1},1}\sqrt{n_{1}m_{1}}%
<n_{1}-1,n_{2}|\rho |m_{1}-1,m_{2}>$

$+\gamma _{R}\delta _{n_{2},0}\delta _{m_{2},0}(-1)^{n_{1}+m_{1}}\sqrt{%
(n_{2}+1)(m_{2}+1)}<n_{1},n_{2}+1|\rho |m_{1},m_{2}+1>$

$-i\Omega \{\delta _{n_{1},1}\delta _{n_{2},0}(-1)^{n_{1}-1}\sqrt{%
n_{1}(n_{2}+1)}<n_{1}-1,n_{2}+1|\rho |m_{1}m_{2}>$

$+\delta _{n_{1},0}\delta _{n_{2},1}(-1)^{n_{1}}\sqrt{(n_{1}+1)n_{2}}%
<n_{1}+1,n_{2}-1|\rho |m_{1}m_{2}>$

$-\delta _{m_{1},0}\delta _{m_{2},1}(-1)^{m_{1}}\sqrt{(m_{1}+1)m_{2}}%
<n_{1},n_{2}|\rho |m_{1}+1,m_{2}-1>$

$-\delta _{m_{1},1}\delta _{m_{2},0}(-1)^{m_{1-1}}\sqrt{(m_{2}+1)m_{1}}%
<n_{1},n_{2}|\rho |m_{1}-1,m_{2}+1>\}$\newline

where $n_1,\ n_2$ refer to the occupation number of the first and second
wells respectively.

Note that the diagonal matrix elements represent the occupation
probabilities of each well, 
\begin{equation}
P(n_{1},n_{2},t)=\langle n_{1},n_{2}|\rho (t)|n_{1},n_{2}\rangle
\end{equation}
The diagonal matrix elements then obey the equation,

$%
{\displaystyle {d \over dt}}%
<n_{1}n_{2}|\rho |n_{1}n_{2}>$

$=[-\gamma _{L}(n_{1}+1)\delta _{n_{1},0}-\gamma _{R}n_{2}\delta
_{n_{2},1}]<n_{1}n_{2}|\rho |n_{1}n_{2}>$

$+\delta _{n_{1},1}\gamma _{L}n_{1}<n_{1}-1,n_{2}|\rho |n_{1}-1,n_{2}>$

$+\delta _{n_{2},0}\gamma _{R}(n_{2}+1)<n_{1},n_{2}+1|\rho |n_{1},n_{2}+1>$

$+i(-1)^{n_{1}}\Omega \{\delta _{n_{1},1}\delta _{n_{2},0}\sqrt{%
n_{1}(n_{2}+1)}[<n_{1}-1,n_{2}+1|\rho |n_{1}n_{2}>-<n_{1},n_{2}|\rho
|n_{1}-1,n_{2}+1>$

$+\delta _{n_{1},0}\delta _{n_{2},1}\sqrt{(n_{1}+1)n_{2}}[<n_{1},n_{2}|\rho
|n_{1}+1,n_{2}-1>-<n_{1}+1,n_{2}-1|\rho |n_{1}n_{2}>]\}$

we now define the off-diagonal matrix elements as\newline

$Y_{1}\equiv <n_{1},n_{2}|\rho |n_{1}+1,n_{2}-1>$

$Y_{2}\equiv <n_{1},n_{2}|\rho |n_{1}-1,n_{2}+1>$

$Y_{3}\equiv <n_{1}-1,n_{2}+1|\rho |n_{1}+1,n_{2}-1>$

Therefore, the population equation we are interested is,\newline

$%
{\displaystyle {d \over dt}}%
P(n_{1},n_{2},t)$

$=[-\gamma _{L}(n_{1}+1)\delta _{n_{1},0}-\gamma _{R}n_{2}\delta
_{n_{2},1})]P(n_{1},n_{2},t)$

$+\delta _{n_{1},1}\gamma _{L}n_{1}P(n_{1}-1,n_{2},t)+\delta
_{n_{2},0}\gamma _{R}(n_{2}+1)P(n_{1},n_{2}+1,t)$

$-2\Omega (-1)^{n_{1}}[\delta _{n_{1},0}\delta _{n_{2},1}\sqrt{(n_{1}+1)n_{2}%
}%
\mathop{\rm Im}%
Y_{1}-\delta _{n_{1},1}\delta _{n_{2},0}\sqrt{n_{1}(n_{2}+1)}%
\mathop{\rm Im}%
Y_{2}]$

Note that the elastic scattering rates, $\eta_1,\eta_2$, do not directly
enter this equation. This is because elastic scattering does not change the
occupation of the well states but does disrupt the phase coherence between
the wave functions in the wells. This will lead to a decay of the relevant
off-diagonal matrix elements, which obey the equations,\newline

$%
{\displaystyle {d \over dt}}%
Y_{1}(t)=%
{\displaystyle {d \over dt}}%
<n_{1}n_{2}|\rho |n_{1}+1,n_{2}-1>$

$=\{-%
{\displaystyle {\gamma _{L} \over 2}}%
(n_{1}+1)\delta _{n_{1},0}-%
{\displaystyle {\gamma _{R} \over 2}}%
n_{2}\delta _{n_{2},1}-\eta _{1}[n_{1}^{2}\delta
_{n_{1},1}+(n_{1}+1)^{2}\delta _{n_{1}+1,1})]-\eta _{2}n_{2}^{2}\delta
_{n_{2},1}\}Y_{1}(t)$

$-i(-1)^{n_{1-1}}\Omega \delta _{n_{1},1}\delta _{n_{2},0}\sqrt{%
n_{1}(n_{2}+1)}Y_{3}(t)$

$-i(-1)^{n_{1}}\Omega \delta _{n_{1},0}\delta _{n_{2},1}\sqrt{(n_{1}+1)n_{2}}%
P(n_{1}+1,n_{2}-1,t)$

$+i(-1)^{n_{1}}\Omega \delta _{n_{1}+1,1}\delta _{n_{2}-1,,0}\sqrt{%
(n_{1}+1)n_{2}}P(n_{1},n_{2},t)$\newline

$%
{\displaystyle {d \over dt}}%
Y_{2}(t)=%
{\displaystyle {d \over dt}}%
<n_{1}n_{2}|\rho |n_{1}-1,n_{2}+1>$

$=\{-%
{\displaystyle {\gamma _{L} \over 2}}%
[(n_{1}+1)\delta _{n_{1},0}+n_{1}\delta _{n_{1}-1,0}]-%
{\displaystyle {\gamma _{R} \over 2}}%
[n_{2}\delta _{n_{2},1}+(n_{2}+1)\delta _{n_{2}+1,1}]$

$-\eta _{1}n_{1}^{2}\delta _{n_{1},1}-\eta _{2}[n_{2}^{2}\delta
_{n_{2},1}+(n_{2}+1)^{2}\delta _{n_{2}+1,1}]\}Y_{2}(t)$

$-i(-1)^{n_{1}}\Omega \delta _{n_{1},0}\delta _{n_{2},1}\sqrt{(n_{1}+1)n_{2}}%
Y_{3}^{*}(t)$

$-i(-1)^{n_{1}-1}\Omega \delta _{n_{1},1}\delta _{n_{2},0}\sqrt{%
n_{1}(n_{2}+1)}P(n_{1}-1,n_{2}+1,t)$

$+i(-1)^{n_{1-1}}\Omega \delta _{n_{1}-1,0}\delta _{n_{2}+1,,1}\sqrt{%
n_{1}(n_{2}+1)}P(n_{1},n_{2},t)$\newline

$%
{\displaystyle {d \over dt}}%
Y_{3}(t)=%
{\displaystyle {d \over dt}}%
<n_{1}-1,n_{2}+1|\rho |n_{1}+1,n_{2}-1>$

$=[-%
{\displaystyle {\gamma _{L} \over 2}}%
n_{1}\delta _{n_{1}-1,0}-%
{\displaystyle {\gamma _{R} \over 2}}%
(n_{2}+1)\delta _{n_{2}+1,1}-\eta _{1}(n_{1}+1)^{2}\delta _{n_{1}+1,1}-\eta
_{2}(n_{2}+1)^{2}\delta _{n_{2}+1,1}]Y_{3}(t)$

$-i(-1)^{n_{1}}\Omega \delta _{n_{1}-1,0}\delta _{n_{2}+1,1}\sqrt{%
n_{1}(n_{2}+1)}Y_{1}(t)+i(-1)^{n_{1}}\Omega \delta _{n_{1}+1,1}\delta
_{n_{2}-1,,0}\sqrt{(n_{1}+1)n_{2}}Y_{2}^{*}(t)$.\newline

To proceed we solve the equations for $Y_1,Y_2,Y_3$ in the steady state,
assuming that the diagonal matrix elements are time constant in time over
the lifetime of the of-diagonal matrix elements. This is the adiabatic
approximation. These steady state values are then substituted back into the
equation for the diagonal matrix elements to obtain a classical jump process
master equation to describe sequential tunnelling. The algebra is tedious,
so we will not give details. The result is\newline

$%
{\displaystyle {d \over dt}}%
P(n_{1},n_{2},t)$

$=-\gamma _{L}[(n_{1}+1)\delta _{n_{1},0}-\gamma _{R}n_{2}\delta
_{n_{2},1})]P(n_{1},n_{2},t)$

$+\delta _{n_{1},1}\gamma _{L}n_{1}P(n_{1}-1,n_{2},t)+\delta
_{n_{2},0}\gamma _{R}(n_{2}+1)P(n_{1},n_{2}+1,t)$

$-2\Omega ^{2}\{\delta _{n_{1},0}\delta _{n_{2},1}(n_{1}+1)n_{2}%
{\displaystyle {(a_{22}a_{33}-a_{23}a_{32}) \over D}}%
[P(n_{1}+1,n_{2}-1,0)-P(n_{1},n_{2},0)]$

$+\delta _{n_{1},1}\delta _{n_{2},0}n_{1}(n_{2}+1)%
{\displaystyle {(a_{11}a_{33}-a_{13}a_{31}) \over D}}%
[P(n_{1}-1,n_{2}+1,0)-P(n_{1},n_{2},0)]\}$.\newline

where $D$ is given by 
\begin{equation}
D=a_{11}(a_{22}a_{33}-a_{23}a_{32})-a_{13}a_{22}a_{31}
\end{equation}
with\newline
$a_{11}=-%
{\displaystyle {\gamma _{L} \over 2}}%
(n_{1}+1)\delta _{n_{1},0}-%
{\displaystyle {\gamma _{R} \over 2}}%
n_{2}\delta _{n_{2},1}-\eta _{1}[n_{1}^{2}\delta
_{n_{1},1}+(n_{1}+1)^{2}\delta _{n_{1},0})-\eta _{2}n_{2}^{2}\delta
_{n_{2},1}$\newline

$a_{13}=(-1)^{n_{1}}\Omega \delta _{n_{1},1}\delta _{n_{2},0}\sqrt{%
n_{1}(n_{2}+1)}=-a_{31}$

$a_{22}=-%
{\displaystyle {\gamma _{L} \over 2}}%
[(n_{1}+1)\delta _{n_{1},0}+n_{1}\delta _{n_{1},1}]-%
{\displaystyle {\gamma _{R} \over 2}}%
[n_{2}\delta _{n_{2},1}+(n_{2}+1)\delta _{n_{2},0}]$\newline

$-\eta _{1}n_{1}^{2}\delta _{n_{1},1}-\eta _{2}[n_{2}^{2}\delta
_{n_{2},1}+(n_{2}+1)^{2}\delta _{n_{2},0}]$\newline

$a_{23}=-(-1)^{n_{1}}\Omega \delta _{n_{1},0}\delta _{n_{2},1}\sqrt{%
(n_{1}+1)n_{2}}=-a_{32}$\newline
$a_{33}=-%
{\displaystyle {\gamma _{L} \over 2}}%
n_{1}\delta _{n_{1,1}}-%
{\displaystyle {\gamma _{R} \over 2}}%
(n_{2}+1)\delta _{n_{2},0}-\eta _{1}(n_{1}+1)^{2}\delta _{n_{1},0}-\eta
_{2}(n_{2}+1)^{2}\delta _{n_{2},0}$\newline

In addition to the incoherent tunnelling of electrons between the wells and
the external reservoirs, we now have incoherent (sequential) tunnelling
between the two wells at rates determined by $\frac{\Omega ^{2}}{\eta _{i}}$%
. The form of this equation corresponds to the sequential tunnelling master
equation obtained by Carlos Egues et al\cite{Egues94}. We have thus shown
that, in the limit of strong decoherence induced by elastic scattering of
the bound states, a population master equation may describe sequential
tunnelling in the device. This will be the appropriate limit in the case
that $\eta _{i}\ >>\ \Omega $. However future quantum nanostructures devices
are likely to operate in the opposite limit. In that case our method is
ideally suited for determining the device characteristics.

\section{Summary}

We have shown how the quantum theory of open systems, formulated as a
quantum stochastic process, enables the current noise spectrum to be
calculated for mesoscopic tunnelling devices. Our approach explicitly treats
quantum noise properties of the charge carriers, and gives a simple
intuitive picture to understand the results. As fabrication technology
develops, quantum noise limited networks of {\it coherent} tunnelling
devices, such as quantum dots and quantum point contacts, will become
increasingly important. Such coherently coupled devices are essential for
the implementation of a quantum computer, which must operate reversibly\cite
{quantumcomp}
. The full operator master equation methods we have demonstrated here
provide a powerful description, including both diagonal and off-diagonal
matrix elements in the same equation.

Our model does not treat the transverse unbound modes in the well of a
realistic resonant tunnelling device. These can easily be incorporated by
additional states in the well and additional jump process channels in the
master equation. We have not done that here as we sought to derive the
irreducible level of current noise in tunnelling devices. Our model may in
fact apply to very tightly confined quantum dot structures which could
conceivably be fabricated with a single bound well states at donor
impurities. Further extensions of the model are also needed to treat the
case where the well state is just below the Fermi level in the collector in
which case the current noise acquire an additional temperature dependent
classical component. These more general cases will be treated in a larger
publication.

\acknowledgments

H.B. SUN would like to thank H. Wiseman for useful discussions.

\appendix

In this appendix we derive the master equation describing elastic scattering
of the quasi-bound states of the well which cause a dephasing of the
electron quasi-bound states but do not change their populations. The
Hamiltonian for double-well system in Schroedinger picture is

\begin{equation}
H=H_{0}+H_{T}+H_{scat}  \eqnum{A1}
\end{equation}

\begin{eqnarray}
H_{0} &=&\sum_{n=1}^{2}\varepsilon _{n}c_{n}^{\dagger
}c_{n}+\sum_{k}\varepsilon _{k}^{E}a_{k}^{\dagger }a_{k}+\sum_{p}\varepsilon
_{p}^{C}b_{p}^{\dagger }b_{p}  \nonumber \\
&&+\sum_{q}\omega _{q}a_{q}^{\dagger }a_{q}+\Omega (c_{1}^{\dagger
}c_{2}+c_{2}^{\dagger }c_{1})  \eqnum{A2}
\end{eqnarray}

\begin{equation}
H_{T}=\sum_{k}(T_{Ek}c^{\dagger }a_{k}+T_{Ek}^{*}a_{k}^{\dagger
}c)+\sum_{p}(T_{Cp}b_{p}^{\dagger }c+T_{Cp}^{*}c^{\dagger }b_{p})  \eqnum{A3}
\end{equation}

\begin{equation}
H_{scat}=\sum_{n=1}^{2}c_{n}^{\dagger }c_{n}\Gamma _{n}  \eqnum{A4}
\end{equation}
where 
\begin{eqnarray}
\Gamma _{1} &=&\sum\limits_{q}M_{q}(\alpha _{q}^{\dagger }+\alpha _{q}) 
\eqnum{A5} \\
\Gamma _{2} &=&\sum\limits_{q}M_{q}(\beta _{q}^{\dagger }+\beta _{q}) 
\eqnum{A6}
\end{eqnarray}
where $\alpha _{q},\beta _{q}$ are Bose destruction operators describing
independent reservoir oscillators. Note that each bound state in the well is
coupled to an independent reservoir. This assumes that there are no
correlations between well states due to the dephasing that takes place
through elastic collisions.

We will only consider here the derivation of the master equation arising
from the elastic scattering of bound states and the harmonic oscillator
reservoirs. The relevant part of the master equation is \cite{Gardiner}

\begin{equation}
\frac{d\rho (t)}{dt}=-\frac{1}{\hbar ^{2}}\int_{0}^{t}d\tau
Tr_{B}[H_{scat}(t),[H_{scat}(\tau ),\rho (\tau )\otimes \rho _{B}]] 
\eqnum{A7}
\end{equation}
where $\rho _{B}$ is the equilibrium state of the bath, and where $Tr_{B}$
means to trace over the bath variables. This equation is based on a second
order expansion in the interaction energy between the reservoir states and
the bound states of the well. We have also assumed that the system and bath
states are decorrelated very rapidly on the time scale of interest in the
system, so that the bath remains close to its equilibrium state. The bath
Hamiltonian is

\begin{equation}
H_{B}=\sum_{q}\omega _{q}a_{q}^{\dagger }a_{q}  \eqnum{A8}
\end{equation}

It will only be necessary to consider one of the bath-well state coupling
terms in the scattering Hamiltonian. The relevant part of the master
equation in the interaction picture is

\begin{eqnarray}
\frac{d\rho (t)}{dt} &=&-\frac{1}{\hbar ^{2}}\int_{0}^{t}d\tau
Tr_{B}\{[c_{n}^{\dagger }c_{n}\sum\limits_{q^{^{\prime }}}M_{q^{^{\prime
}}}(\alpha _{q^{^{\prime }}}^{\dagger }e^{i\omega _{q^{^{\prime }}}t}+\alpha
_{q^{^{\prime }}}e^{-i\omega _{q^{^{\prime }}}t}),  \nonumber \\
&&[c_{n}^{\dagger }c_{n}\sum\limits_{q}M_{q}(\alpha _{q}^{\dagger
}e^{i\omega _{q}\tau }+\alpha _{q}e^{-i\omega _{q}\tau }),\rho (\tau
)\otimes \rho _{B}]]\}  \eqnum{A9}
\end{eqnarray}

We now define,

\begin{equation}
E\equiv c_{n}^{\dagger }c_{n}\sum\limits_{q^{^{\prime }}}M_{q^{^{\prime
}}}(\alpha _{q^{^{\prime }}}^{\dagger }e^{i\omega _{q^{^{\prime }}}t}+\alpha
_{q^{^{\prime }}}e^{-i\omega _{q^{^{\prime }}}t})  \eqnum{A10}
\end{equation}

\begin{equation}
F\equiv c_{n}^{\dagger }c_{n}\sum\limits_{q}M_{q}(\alpha _{q}^{\dagger
}e^{i\omega _{q}(t-\tau )}+\alpha _{q}e^{-i\omega _{q}(t-\tau )}) 
\eqnum{A11}
\end{equation}

Therefore

\begin{eqnarray}
\frac{d\rho (t)}{dt} &=&-\frac{1}{\hbar ^{2}}\int_{0}^{t}d\tau
Tr_{B}\{EF\rho (\tau )\otimes \rho _{B}-E\rho (\tau )\otimes \rho _{B}F 
\nonumber \\
&&-F\rho (\tau )\otimes \rho _{B}E+\rho (\tau )\otimes \rho _{B}FE\} 
\eqnum{A12}
\end{eqnarray}

The state of the reservoirs is taken to be a thermal state at temperature $T$%
, thus

\begin{eqnarray}
Tr_{B}[\alpha _{q}\alpha _{q^{^{\prime }}}\rho _{B}] &=&Tr_{B}[\alpha
_{q}^{\dagger }\alpha _{q^{^{\prime }}}^{\dagger }\rho _{B}]=0  \eqnum{A13}
\\
Tr_{B}[\alpha _{q}^{\dagger }\alpha _{q^{^{\prime }}}\rho _{B}] &=&\delta
_{qq^{^{\prime }}}%
{\displaystyle {1 \over e^{E_{q}/k_{B}T}-1}}%
\eqnum{A14}
\end{eqnarray}

The first term on the right hand of the equation (A12)

\begin{eqnarray}
&&\int_{0}^{t}d\tau Tr_{B}EF\rho (t)\otimes \rho _{B}  \nonumber \\
&=&\int_{0}^{t}d\tau \sum\limits_{qq^{^{\prime }}}M_{q}M_{q^{^{\prime
}}}\delta _{qq^{^{\prime }}}[%
{\displaystyle {1 \over e^{E_{q}/k_{B}T}-1}}%
e^{i[(\omega _{q^{^{\prime }}}-\omega _{q})t+\omega _{q}\tau ]}  \nonumber \\
&&+(1+%
{\displaystyle {1 \over e^{E_{q}/k_{B}T}-1}}%
e^{i[(\omega _{q}-\omega _{q^{^{\prime }}})t-\omega _{q}\tau
]}](c_{n}^{\dagger }c_{n})^{2}\rho  \nonumber \\
&=&\int_{0}^{t}d\tau \sum\limits_{q}|M_{q}|^{2}[%
{\displaystyle {1 \over e^{E_{q}/k_{B}T}-1}}%
e^{i\omega _{q}\tau }+(1+%
{\displaystyle {1 \over e^{E_{q}/k_{B}T}-1}}%
)e^{-i\omega _{q}\tau }](c_{n}^{\dagger }c_{n})^{2}\rho  \nonumber \\
&=&\sum\limits_{q}|M_{q}|^{2}[%
{\displaystyle {\sin (\omega _{q}t) \over \omega _{q}}}%
(1+%
{\displaystyle {2 \over e^{E_{q}/k_{B}T}-1}}%
)+i%
{\displaystyle {\cos (\omega _{q}t)-1 \over \omega _{q}}}%
](c_{n}^{\dagger }c_{n})^{2}\rho  \nonumber \\
&=&(\eta _{n}+i\xi _{n})(c_{n}^{\dagger }c_{n})^{2}\rho  \eqnum{A15}
\end{eqnarray}

Where

\begin{equation}
\eta _{n}\equiv \sum\limits_{q^{^{\prime }}}|M_{q}|^{2}%
{\displaystyle {\sin (\omega _{q}t) \over \omega _{q}}}%
(1+%
{\displaystyle {2 \over e^{E_{q}/k_{B}T}-1}}%
)  \eqnum{A16}
\end{equation}

\begin{equation}
\xi _{n}\equiv \sum\limits_{q^{^{\prime }}}|M_{q}|^{2}%
{\displaystyle {\cos (\omega _{q}t)-1 \over \omega _{q}}}%
\eqnum{A17}
\end{equation}

Similarly,

\begin{eqnarray}
&&-\int_{0}^{t}d\tau Tr_{B}E\rho (\tau )\otimes \rho _{B}F  \nonumber \\
&=&-(\eta _{n}-i\xi _{n})c_{n}^{\dagger }c_{n}\rho c_{n}^{\dagger }c_{n} 
\eqnum{A18}
\end{eqnarray}

\begin{eqnarray}
&&-\int_{0}^{t}d\tau Tr_{B}F\rho (\tau )\otimes \rho _{B}E  \nonumber \\
&=&-(\eta _{n}+i\xi _{n})c_{n}^{\dagger }c_{n}\rho c_{n}^{\dagger }c_{n} 
\eqnum{A19}
\end{eqnarray}
\begin{eqnarray}
&&\int_{0}^{t}d\tau Tr_{B}\rho (\tau )\otimes \rho _{B}FE  \nonumber \\
&=&(\eta _{n}-i\xi _{n})\rho (c_{n}^{\dagger }c_{n})^{2}  \eqnum{A20}
\end{eqnarray}

The coefficients $\eta _{n},\xi _{n}$ appear to be time dependant, but under
reasonable physical assumptions are time independant\cite{Gardiner}. These
assumptions are, firstly that $t$ is assumed to be a time scale over which
the system operators vary significantly. On this time scale bath correlation
functions decay rapidly. Secondly, that the bath is at finite temperature
and there is significant excitation above the reservoir ground state.
Finally that the coupling constants $M_{q}$ are independent of $q$ up to
some large cut-off wave number. Under these assumptions these coefficients
can be evaluated in the limit of $t\rightarrow \infty $. We refer the reader
to reference \cite{Gardiner} for more details.

The total contribution from the scattering term to the master equation is
therefore by substituting equations (A15) and (A18 -A20) and coresponding
termes for the second well into equation (A12 ):

\begin{equation}
\frac{d\rho (t)}{dt}=\sum_{n=1}^{2}\eta _{n}[c_{n}^{\dagger
}c_{n},[c_{n}^{\dagger }c_{n},\rho ]]+%
{\displaystyle {i\xi  \over \hbar ^{2}}}%
\sum_{n=1}^{2}[c_{n}^{\dagger }c_{n},\rho ]  \eqnum{A21}
\end{equation}

The effect of the term i$\xi $ is to add a small perturbation to the energy
of each quasi-bound well state and is equivalent to the Lamb shift term in
atomic physics.

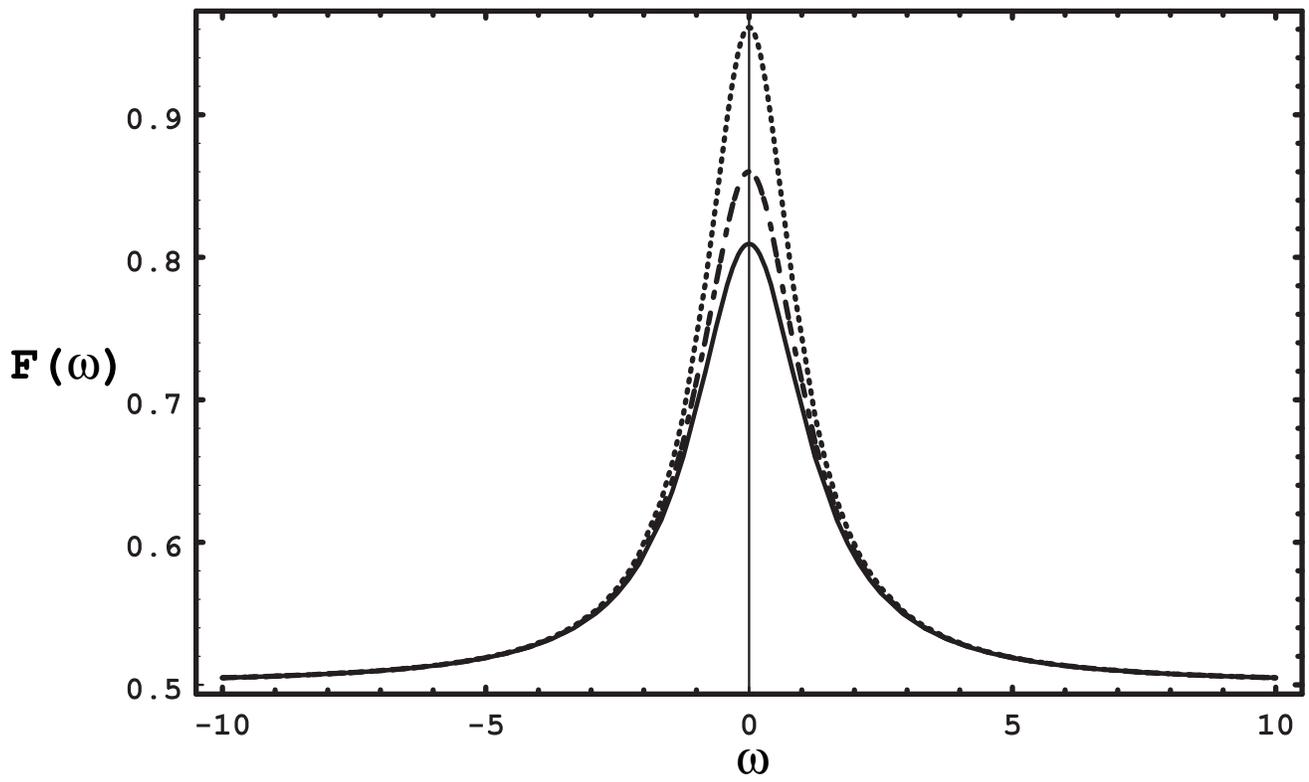
\begin{figure}[tbp]
\caption{The current Fano factor $\frac{S(\omega)}{2ei_\infty}$ versus
normalised frequency $\omega/\gamma_e$ in case 1. All parameters are
normalised by $\gamma_e$ . The corresponding parameters ($\eta_e/\gamma_e$, $%
\omega/\gamma_e$ ) for the curves are from the top: Dotted:(0.5, 0.1);
Dot-dash:(0.5, 0.2); Solid:(0.5, 0.24).}
\label{fig1}
\end{figure}

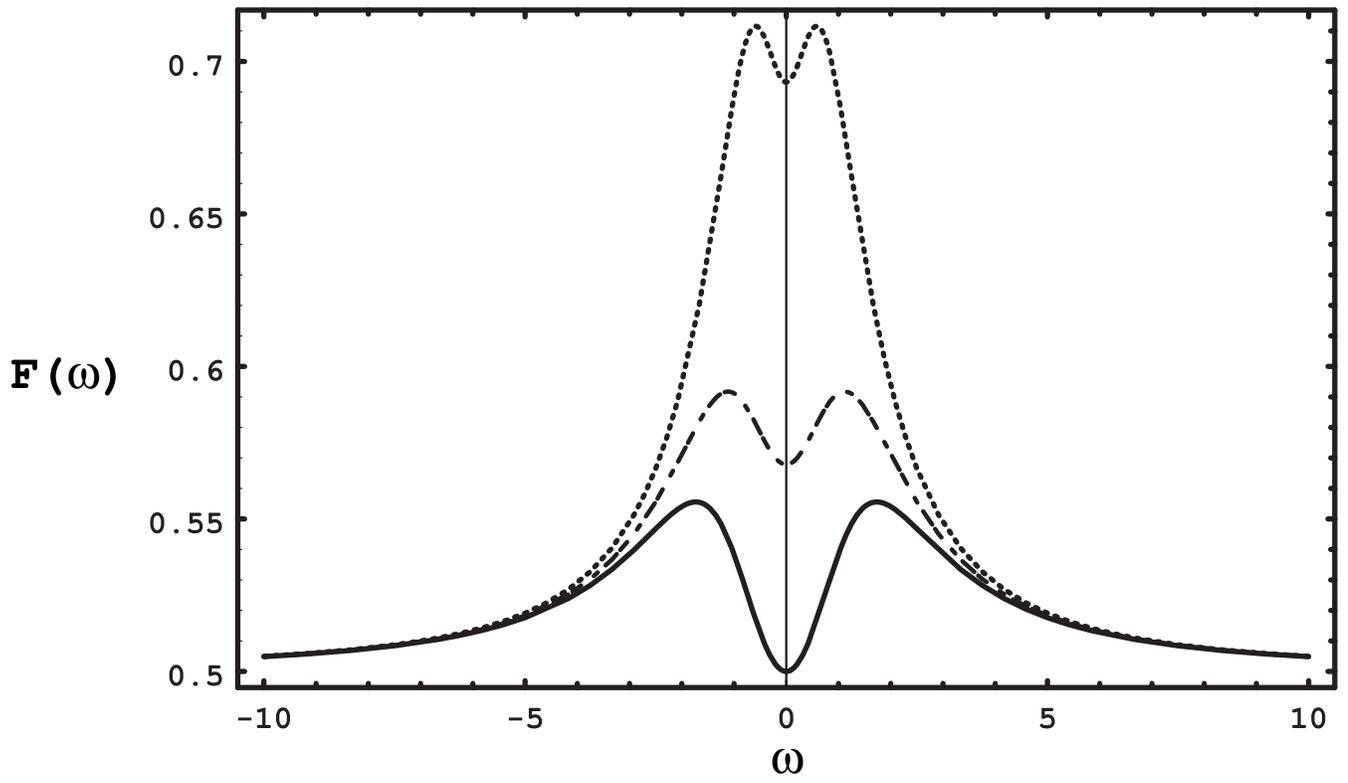
\begin{figure}[tbp]
\caption{The current Fano factor versus normalised frequency ($\omega/
\gamma_e$ in case 2. All parameters are normalised by $\gamma_e$. The
corresponding parameters $\eta_e/\gamma_e$, $\Omega/\gamma_e$) for the
curves are from the top: Dotted:(0, 0.2); Dot-dash:(0.4, 0.4); Solid:(0.5,
0.5).}
\label{fig2}
\end{figure}

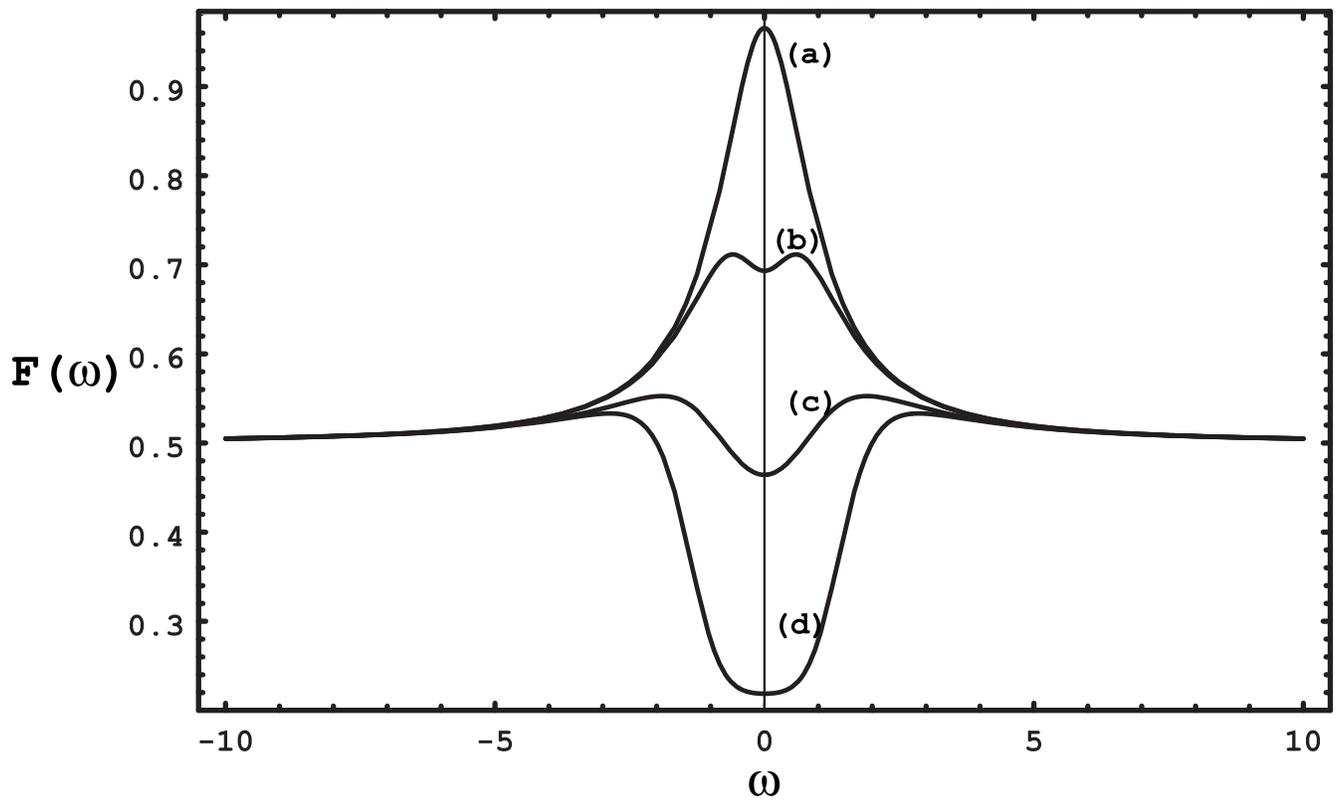
\begin{figure}[tbp]
\caption{ The comparison of influences of the elastic scattering and the
coherent tunnelling. The corresponding parameters ($\eta_e/\gamma_e$, $%
\Omega/\gamma_e$ ) for the curves are from the top: (a):(0.6, 0.1);
(b):(0.0, 0.2); (c):(0.2, 0.5); (d):(0, 0.645).}
\label{fig3}
\end{figure}

\begin{figure}[tbp]
\caption{ The current Fano-factors versus normalized frequency for double
well structures. The normalized parameter $\Omega /\gamma _{e}$ for the
curves are from top: 0.2 dash line , 0.645 dot-dash line, 5.0 solid line.}
\label{fig4}
\end{figure}

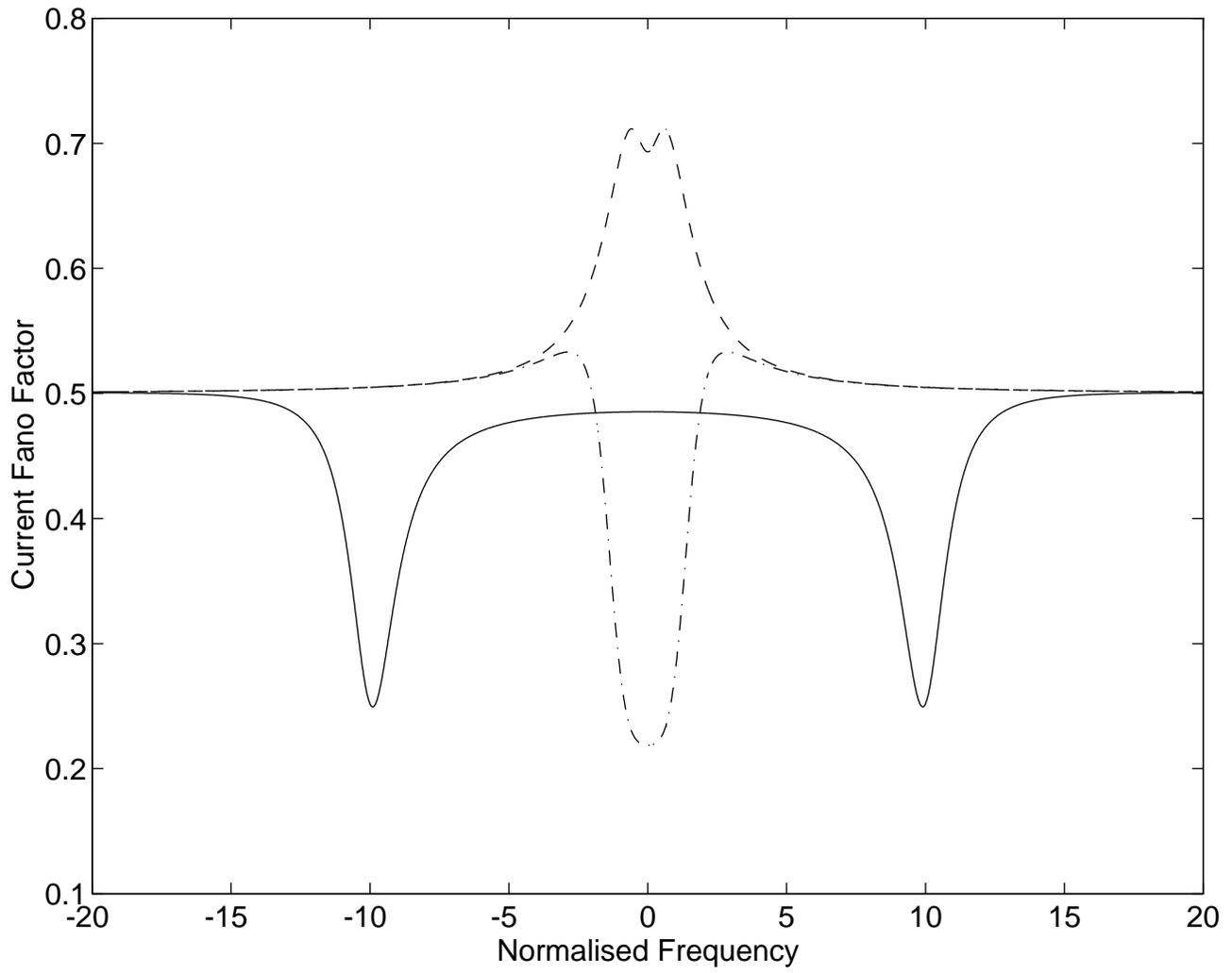
\end{document}